# A note on the separability index

Linda Mthembu, Tshilidzi Marwala

Department of Electrical and Information Engineering, University of the Witwatersrand
Johannesburg, South Africa
linda.mthembu@student.wits.ac.za, t.marwala@ee.wits.ac.za

**Abstract**

In discriminating between objects from different classes, the more separable these classes are the less computationally expensive and complex a classifier can be used. One thus seeks a measure that can quickly capture this separability concept between classes whilst having an intuitive interpretation on what it is quantifying. A previously proposed separability measure, the separability index (SI) has been shown to intuitively capture the class separability property very well. This short note highlights the limitations of this measure and proposes a slight variation to it by combining it with another form of separability measure that captures a quantity not covered by the Separability Index.

## 1. Introduction

In object categorization/classification one is given a dataset of objects from different classes from which to discover a class-distinguishing-pattern so as to predict the classification of new, previously unseen objects [1,7]. This will only be possible if the main justification pillar of induction systems which is based on the dictum; "similar objects tend to cluster together" is true. This process of discovering a pattern in the dataset is further complicated by the fact that the dataset often cannot immediately be visualized to determine the class distribution. This could be due to the datasets' high dimensionality. Discovering a method that can distil such information, *without* running multiple sets of computationally expensive classifiers, would be advantageous.

This method should quantify how the classes are distributed with respect to each other; are there class overlaps, are there multiple modes within the classes and are there many outliers etc? We thus seek a simple measure that can concisely capture some of these aspects of the classes to gauge the complexity of classifier to be implemented. The notion of a 'simpler classifier' relates to the complexity of the discrimination function. A simpler function e.g. linear is preferred over a more complex polynomial function as stated by Occam's razor. The complexity of a classifier is also determined by the number of irrelevant features in the dataset. The original dataset input space – defined by the number of expertly measured attributes - is often not the optimal in terms of producing clearly separable/non-overlapping classes. A subset of this space can often produce a substantially separable set of classes which in turn results in a simpler discriminating function. Searching for an optimal sub-space can be considered an optimization problem whose criterion function is the maximization of some predefined separability measure. A recent review and comment on this area of research is presented in [4 and 6]. One measure, the separability index (SI), that intuitively measures the class overlap was previously introduced in [3, 8] and was shown to be efficient in a number of popular machine learning datasets in [3, 5].

The separability index measure estimates the average number of instances in a dataset that have a nearest neighbour with the same label. Since this is a fraction the index varies between 0-1 or 0-100%. Another separability measure, based on the class distance or margin is the Hypothesis margin (HM), introduced in [2]. It measures the distance between an object's nearest neighbour of the same class (near-hit) and a nearest neighbour of the opposing class (near-miss) and sums over these. This means the larger the near-miss distance and smaller the near-hit values, the larger the hypothesis margin will be.

This note is only concerned with the above two mentioned measures' limitations. In the next section we show with a simple example the behaviour of both the SI and HM. We highlight the advantages and disadvantages of SI and HM then we propose a hybrid of the two measures. The resulting measures' pseudo code and behaviour are presented.

## 2. Separability

### 2.1 Behaviour of separability measures

In this section the behaviour of both measures is simulated in an example where the separation of two Gaussian clusters is incrementally increased. This is taken to simulate the process of searching for an optimal feature space in a given high dimensional dataset. Figure 1 shows two Gaussian clusters that are initially overlapping with a SI of 0.54 or 54%.

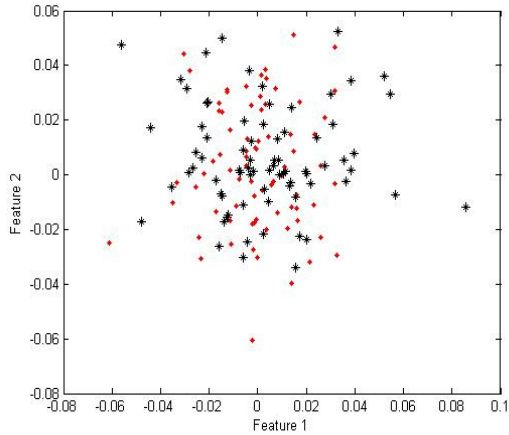

Figure 1: Two initially overlapping classes

These clusters are incrementally separated, by varying one cluster's centre distance from the other. Figure 2 shows the point where the SI measure is 1or 100%; a quadratic or cubic discriminator will certainly be enough to *cleanly* partition the clusters whereas a linear classifier might not without misclassification. Figure 3 shows a state where the two clusters are visually more fully separated than in figure 2 and certainly a linear function will be an adequate classifier for such class separations. Figure 4 shows the variation of the separability index with the increasing cluster distance.

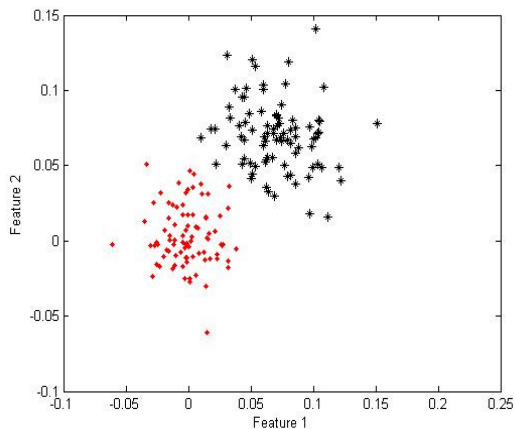

Figure 2: The Separability index is 100%

When the class separation distance increases beyond 0.015 units the SI still reports a separability of 1. It is clear from this figure that the SI is limited in capturing extreme class separability information which could result when a feature sub-space with fewer features than that at 100% separability is discovered in the optimization.

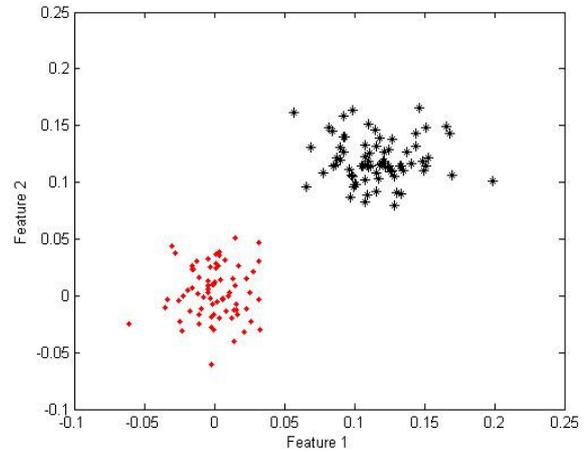

Figure 3: Increased class separability

The SI measure is informative about the separability of the clusters below full separability (<=1) but is no longer informative when the classes separate further which can arise in practise. This is to be expected since the separability index does not measure class distances per se. The hypothesis margin on the other hand, shown in figure 5, keeps on measuring with no real informative limit on the quantity it is measuring except that the class separation distance is increasing. What is required is a measure that has the ability to *intuitively* inform on the class separability below 100%, a characteristic of the separability index and has the ability to continue measuring after 100% class separability, a characteristic of the hypothesis margin.

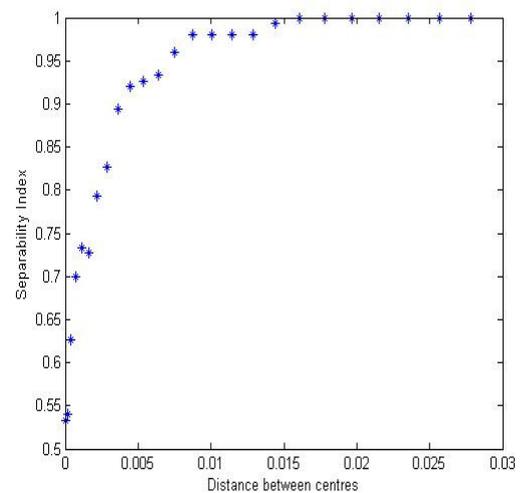

Figure 4: Separability index results on the two Gaussian clusters as the centre distance is increased

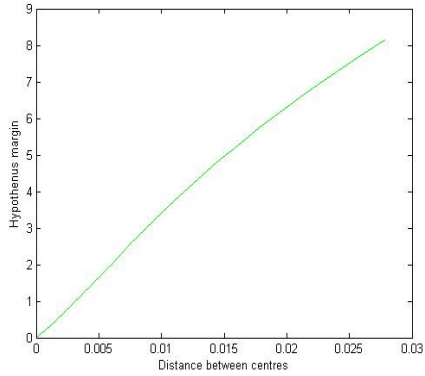

Figure 5: Hypothesis margin results on the two Gaussian clusters as the centre distance is increased

### 3. The Hybrid separability measure

Merging the two measures will consist of two parts; the original SI and *modified* HM parts. The HM is modified by only initializing it when the separability index measures a separability of 1. While the SI is below 1 the HM is set to zero and once the SI is equal to 1 the HM is activated. Subsequent hypothesis margin distances are then calculated as ratios with respect to the HM when the SI was 1.

In this hybrid measure the SI part will capture all the sub-spaces, from feature selection, where the class separability increases until unity then the modified HM part will capture the fact that the clusters are still separating further. This way the hybrid separability measure captures the overall class separability in terms of distance and instance overlap. Figure 5 shows the pseudo code for the proposed algorithm:

```
hm = hypothesis margin; % original hypothesis margin
si  = separability index;  % separability index
if si < 1
   hybrid = 100*Si;   % hybrid measure equal SI when
                     % SI is less than 1.
   hm_ratio =0;      % hypothesis ratio
   hm = 0;           % hypothesis margin
   counter =0;
elseif si = 1
   counter =counter +1;

   if counter =1   % first time SI is 1 capture the
     ih = hm;      % hypothesis margin distance to be
            % the reference for subsequent distances
   end
  hm_ratio = hm/ih;     % hypothesis ratio
  hybrid = 100*hm_ratio; % hybrid measure
end
```

**Figure 5**: Pseudo code for hybrid measure

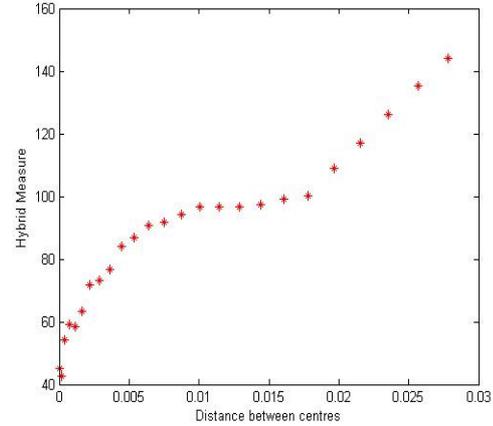

Figure 6: Hybrid measure on the two Gaussian clusters as the centre distance is increased

Figure 6 shows the behaviour of the hybrid separability measure. The SI part is still retained and now the HM part is incorporated as a fraction which is converted to a percentage so as to integrate with the SI measure. The hypothesis margin is now a more informative measure of the class separation. Table 1 below presents a portion of the above simulation results. After the separability index reaches 1 the hypothesis ratio information is relayed to the hybrid measure by multiplying by 100. The separation distance can still be extracted from the hybrid measure.

| SI | HM | HM RATIO | Hybrid |
|---|---|---|---|
| | | | (%) |
| 0.908 | 1.5431 | 0 | 90.8046 |
| 0.9368 | 1.962 | 0 | 93.6782 |
| 0.954 | 2.4002 | 0 | 95.4023 |
| 0.9598 | 2.8622 | 0 | 95.977 |
| 0.9828 | 3.3595 | 0 | 98.2759 |
| 0.9885 | 3.8828 | 0 | 98.8506 |
| 1 | 4.4158 | 1 | 100 |
| 1 | 4.952 | 1.1214 | 112.1431 |
| 1 | 5.4955 | 1.2445 | 124.4502 |
| 1 | 6.0419 | 1.3682 | 136.8238 |
| 1 | 6.5924 | 1.4929 | 149.2898 |
| 1 | 7.1469 | 1.6185 | 161.8487 |
| 1 | 7.7037 | 1.7446 | 174.457 |
| 1 | 8.2627 | 1.8712 | 187.1161 |

Table 1: A sub-set of the simulation results

Intuitive interpretation, in the new measure, is not completely lost and can be derived from the last two columns of table 1. Once the hybrid measure reports separabilities of more than 100% then a different perspective on separability can be induced; the reported quantity will then be the percentage ratio of the class separability distances. A value of 124% can be read to mean the classes are one point two four *times further* apart than they were when the SI index was 100%.

This retains the intuitive notion of average distance between classes (measured by the hypothesis margin (HM)) albeit it is measured from a different reference point, the point at which the separability index (SI) measures 100%.

## 4. Conclusion

This note highlights the advantages and disadvantages of two previously proposed separability measures; the separability index and the hypothesis margin. A hybrid measure is formed from the two and the good properties of the individual measures are retained in the new measure which overcomes the limitations of the previous measures. A simple simulation example exposes the problem of the two measures and performance results of the new measure are presented on the same example. Some intuitive interpretation can still be developed from the new measure.

## 5. Acknowledgements

This research was supported by the financial assistance of the National Research Foundation of South Africa.

## 6. References

[1] R.O. Duda, P. E. Hart and D.G. Stork. Pattern Classification (2nd edition) John Wiley and Sons, 2000.

[2] R. Gilad-Bachrach, A. Navot and N. Tishby. Margin based feature selection- Theory and algorithms. Proc. 21$^{st}$ International Conference on Machine Learning (ICML), Banff, Canada 2004.

[3] Greene J.R. Feature subset selection using Thornton's separability index and its applicability to a number of sparse proximity-based classifiers. In proceedings of the Pattern Recognition Association of South Africa, 2001.

[4] Guyon I and Elisseeff A. An Introduction to Variable and Feature Selection. Journal of Machine Learning Research 3 pages 1157-1182 (2003).

[5] L. Mthembu and J.R Greene. A comparison of three separability measures. In Proc of the 15$^{th}$ Annual symposium of the Pattern Recognition Association of South Africa (PRASA), November 2004, Grabouw, South Africa.

[6] A. Navot, R. Gilad-Bachrach, Y. Navot and N. Tishby. Is feature selection still necessary? In Saunders C and Grobelnik M and Gunn S and Shawe-Taylor J. Editors Latent structure and feature selection techniques: Statistical and Optimisation perspectives workshop (2006).

[7] T. Mitchell. Machine Learning. Published by McGraw Hill, 1997, ISBN 0070428077.

[8] C. Thornton. Truth from Trash: How Learning Makes Sense. Published by MIT Press, 2002, ISBN 0262700875, 9780262700870.